\documentclass[a4paper,10pt]{article}

\usepackage{epsf}
\usepackage{graphicx}    
\usepackage{dcolumn}    
\usepackage{bm}        

\newcommand{\be}[1]{\begin{equation}\label{#1}}
\newcommand{\ee}{\end{equation}}
\newcommand{\bea}{\begin{eqnarray}}
\newcommand{\eea}{\end{eqnarray}}
\def\disp{\displaystyle}

\renewcommand{\markright}{\markright{\thepage}}

\begin{document}

\begin{titlepage}

\begin{flushright}
hep-th/0501160
\end{flushright}

\vspace{5mm}
\begin{center}

{\Large \bf Hessence: A New View of Quintom Dark Energy}

\vspace{15mm}

{\large \bf Hao Wei$\,^{a,b,}$\footnote{Email address:
haowei@itp.ac.cn}, Rong-Gen Cai$\,^{a,c,}$\footnote{Email address:
cairg@itp.ac.cn},
 Ding-Fang Zeng$\,^{a,b,}$\footnote{Email address:
 dfzeng@itp.ac.cn} }\\

\vspace{5mm}
 {\em $^a$Institute of Theoretical Physics, Chinese Academy of Sciences, \\
P.O. Box 2735, Beijing 100080, China \\
 $^b$Graduate School of the Chinese Academy of Sciences, Beijing 100039, China\\
 $^c$CASPER, Department of Physics, Baylor University, Waco, TX76798-7316, USA}

\end{center}

\vspace{12mm}
\begin{abstract}
Recently a lot of attention has been drawn to build dark energy
model in which the equation-of-state parameter $w$ can cross the
phantom divide $w=-1$. One of models to realize crossing the
phantom divide is called quintom model, in which two real scalar
fields appears, one is a normal scalar field and the other is a
phantom-type scalar field. In this paper we propose a
non-canonical complex scalar field as the dark energy, which we
dub ``hessence'', to implement crossing the phantom divide, in a
similar sense as the quintom dark energy model. In the hessence
model, the dark energy is described by a single field with an
internal degree of freedom rather than two independent real scalar
fields. However, the hessence is different from an ordinary
complex scalar field, we show that the hessence can avoid the
difficulty of the Q-ball formation which gives trouble to the
spintessence model (An ordinary complex scalar field acts as the
dark energy). Furthermore, we find that, by choosing a proper
potential, the hessence could correspond to a Chaplygin gas at
late times.
\end{abstract}

\end{titlepage}

\newpage

\setcounter{page}{1}


\section{Introduction}
A lot of cosmological observations, such as SNe Ia~\cite{r1},
WMAP~\cite{r2}, SDSS~\cite{r3}, Chandra X-ray
Observatory~\cite{r4} etc., reveal some cross-checked information
of our universe. They suggest that the universe is spatially flat,
and consists of approximately $70\%$ dark energy with negative
pressure, $30\%$ dust matter (cold dark matters plus baryons), and
negligible radiation, and that the universe is undergoing an
accelerated expansion.

To accelerate the expansion, the equation-of-state parameter
$w\equiv p/\rho$ of the dark energy must satisfy $w<-1/3$, where
$p$ and $\rho$ are its pressure and energy density, respectively.
The simplest candidate of the dark energy is a tiny positive
time-independent cosmological constant $\Lambda$, for which
$w=-1$. Another possibility is quintessence~\cite{r5,r34}, a cosmic
real scalar field that is displaced from the minimum of its
potential. With the evolution of the universe, the scalar field
slowly rolls down its potential. To be definite, we consider the
action\footnote{We adopt the metric convention as $(+,-,-,-)$
throughout this paper.} \be{eq1} S=\int
d^{4}x\sqrt{-g}\left(-\frac{\cal R}{16\pi G}+{\cal L}_{DE}+{\cal
L}_m\right), \ee where $g$ is the determinant of the metric
$g_{\mu\nu}$, $\cal R$ is the Ricci scalar, ${\cal L}_{DE}$ and
${\cal L}_m$ are the Lagrangian densities of the dark energy and
matter, respectively. The Lagrangian density for the quintessence
is
 \be{eq2}
 {\cal
L}_{DE}=\frac{1}{2}\left(\partial_{\mu}\varphi\right)^2-V(\varphi),
\ee where $\varphi$ is a real scalar field. Considering a
spatially flat FRW universe and assuming that the scalar field
$\varphi$ is homogeneous, one has the equation-of-state parameter
of quintessence as
\be{eq3}
w=\frac{\dot{\varphi}^{2}/2-V(\varphi)}{\dot{\varphi}^{2}/2+V(\varphi)}.
\ee
 It is easy to see that $-1\leq w \leq +1$ for quintessence. On
the other hand, the observations cannot exclude the possibility of
phantom matter with $w <-1$~\cite{r6,r7,r33}.  One of ways to realize
the phantom matter is a scalar field with a ``wrong'' sign kinetic
energy term. The Lagrangian density for the phantom scalar field
is given by
 \be{eq4}
 {\cal
L}_{DE}=-\frac{1}{2}\left(\partial_{\mu}\varphi\right)^2-V(\varphi).
\ee
 Its equation-of-state parameter
 \be{eq5}
w=\frac{-\dot{\varphi}^{2}/2-V(\varphi)}{-\dot{\varphi}^{2}/2+V(\varphi)},
 \ee
 clearly, one has  $w\leq-1$ with $\rho = -\dot{\varphi}^{2}/2+V(\varphi)>0$.

Actually, by fitting the recent SNe Ia data, marginal ($2\sigma$)
evidence for $w(z)<-1$ at $z<0.2$ has been found~\cite{r8}. In
addition, many best fit value of $w_0$ are less than $-1$ in
various data fittings with different parameterizations
(see~\cite{r9} for a recent review). The present data seem to
favor an evolving dark energy with $w$ being below $-1$ around
present epoch from $w>-1$ in the near past~\cite{r10}. Obviously,
the equation-of-state parameter $w$ cannot cross the phantom
divide $w=-1$ for quintessence or phantom alone. Recently,  some
efforts have been made to build dark energy model whose the
equation-of-state parameter can cross the divide $w=-1$. Although
some variants of the k-essence~\cite{r11} look possible to give
promising solutions, a no-go theorem, shown in~\cite{r12},
shatters this kind of hopes: It is impossible to cross the phantom
divide $w=-1$, provided that the following conditions are
satisfied: (1) classical level, (2) GR is valid, (3) single real
scalar field, (4) arbitrary Lagrangian density $p\,(\varphi,X)$,
where $X\equiv\frac{1}{2}
g^{\mu\nu}\partial_{\mu}\varphi\partial_{\nu}\varphi$ is the
kinetic energy term, and (5) $p\,(\varphi,X)$ is continuous
function and is differentiable enoughly. Thus, to implement the
transition from $w>-1$ to $w<-1$ or vice versa, it is necessary to
give up at least one of conditions mentioned above.

Obviously, the simplest way to get around this no-go theorem is to
consider a two real scalar field model, i.e. to break the third
condition. In~\cite{r13}, Hu considered a phenomenological model
with two real scalar fields (see also~\cite{r21}) and showed that
it is possible to cross the phantom divide $w=-1$.  Feng, Wang and
Zhang in \cite{r10} proposed a so-called quintom model which is a
hybrid of quintessence and phantom (thus the name quintom).
Naively, one may consider a Lagrangian density~\cite{r10,r14}
\be{eq6} {\cal
L}_{DE}=\frac{1}{2}\left(\partial_{\mu}\phi_1\right)^2-\frac{1}{2}\left(\partial_{\mu}
\phi_2\right)^2-V(\phi_1,\phi_2), \ee where $\phi_1$ and $\phi_2$
are two real scalar fields and play the roles of quintessence and
phantom respectively. Considering a spatially flat FRW universe
and assuming the scalar fields $\phi_1$ and $\phi_2$ are
homogeneous, one obtains the {\em effective} pressure and energy
density for the quintom \bea
p=\frac{1}{2}\dot{\phi}_{1}^2-\frac{1}{2}\dot{\phi}_{2}^2-V(\phi_1,\phi_2),\label{eq7}\\
\rho=\frac{1}{2}\dot{\phi}_{1}^2-\frac{1}{2}\dot{\phi}_{2}^2+V(\phi_1,\phi_2).\label{eq8}
\eea And then, the corresponding {\em effective} equation-of-state
parameter is given by \be{eq9}
w=\frac{\dot{\phi}_{1}^2-\dot{\phi}_{2}^2-2V(\phi_1,\phi_2)}{\dot{\phi}_{1}^2
-\dot{\phi}_{2}^2+2V(\phi_1,\phi_2)}. \ee It is easy to see that
$w\geq-1$ when $\dot{\phi}_{1}^2\geq\dot{\phi}_{2}^2$ while $w<-1$
when $\dot{\phi}_{1}^2<\dot{\phi}_{2}^2$. The cosmological
evolution of the quintom model without direct coupling between
$\phi_1$ and $\phi_2$ was studied by Guo {\it et al.}~\cite{r14}.
They showed that the transition from $w>-1$ to $w<-1$ or vice
versa is possible in this type of quintom model.

In many of the existing quintom-type
models~\cite{r10,r14,r13,r21,r22,r35}, they invoke two independent
real scalar fields to describe the dark energy. However,  it is
also natural to consider a {\em single} field with an {\em
internal degree of freedom} to describe the dark energy. For the
spintessence model of dark energy~\cite{r15,r16,r17,r18} with a
single complex scalar field, it suffers from the problem of Q-ball
formation~\cite{r17,r15,r19}. For the so-called $SO(1,\eta)$ model
of dark energy~\cite{r23}, Wei {\it et al.} extended the unit
imaginary number $i$ to a new parameter $i_\eta$ and constructed
an extended complex scalar field as dark energy.

In fact, by a new view of the quintom model, we propose a
non-canonical complex scalar field, which we dub ``hessence'', to
play the role of quintom. The hessence is similar to the extended
complex scalar field proposed in \cite{r23} in some sense.
However, the motivation and emphasis here are  different from
those in \cite{r23}. The hessence could be viewed as a new window
to look into the unknown internal world of the mysterious dark
energy. In addition, like the case of canonical complex scalar
field, the hessence has a {\em conserved charge}. In section 2 we
will discuss some aspects of the hessence model. In section 3, we
will show that, different from the case of ordinary complex scalar
field,  the hessence can avoid the difficulty of Q-ball formation
which gives trouble to the spintessence. (Q-ball is a kind of
nontopological soliton whose stability is guaranteed by some
conserved charge.) In section 4, we show that, by choosing a
proper potential, the hessence could correspond to a Chaplygin gas
at late times. A brief summary and some discussions are presented
in section 5.


\section{Hessence}

\subsection{Motivation}

Consider a non-canonical complex scalar field as the dark energy,
\be{eq10} \Phi=\phi_1+i\phi_2, \ee with a Lagrangian density
\be{eq11} {\cal L}_{DE}=\frac{1}{4}\left[(\partial_\mu
\Phi)^2+(\partial_\mu \Phi^\ast)^2\right] -V(\Phi,\Phi^\ast). \ee
Obviously, this Lagrangian density is identified with
Eq.~(\ref{eq6}) in terms of two real scalar fields $\phi_1$ and
$\phi_2$. By this formalism, however, the dark energy is described
by a single field rather than two independent fields. The physical
content is changed. On the other hand, this Lagrangian density of
the hessence is different from the case of a canonical complex
scalar field $\Psi$ whose Lagrangian density is given by \be{eq12}
{\cal
L}_{DE}=\frac{1}{2}\left(\partial^\mu\Psi^\ast\right)\left(\partial_\mu\Psi\right)-V(|\Psi|),
\ee where $|\Psi|$ is the absolute value of $\Psi$, namely
$|\Psi|^2=\Psi^\ast\Psi$. Thus, we give this non-canonical complex
scalar field a new name ``hessence'' (the meaning of this name
will be clear below) to make a distinction with the canonical
complex scalar field. In fact, the canonical complex scalar field
was considered as a variant of quintessence for several years,
while it was dubed ``spintessence''~\cite{r15,r16,r17,r18}. The
spintessence also has a conserved charge. However, it is
overlooked as a viable candidate of dark energy because it is
troubled by the Q-ball formation~\cite{r17,r15,r19} which we will
discuss in section 3. We find that it is suggestive to compare the
hessence with the spintessence since they are similar in many
aspects. (Of course, they also have many differences which are
crucial to make the hessence avoid the difficulty of Q-ball
formation.)

The most interesting feature of a complex scalar field different
from a real scalar field is that the complex scalar field has a
conserved charge due to internal symmetry. It is suggestive to
review the case of canonical complex scalar field at first. In
terms of $\Psi=\psi_1+i\psi_2$, the Lagrangian density of
canonical complex scalar field, Eq.~(\ref{eq12}), becomes
$${\cal L}_{DE}=\frac{1}{2}(\partial_\mu\psi_1)^2+
\frac{1}{2}(\partial_\mu\psi_2)^2-V(|\Psi|).$$
It is invariant under the transformation
$$\psi_1\to\psi_1\cos\alpha-\psi_2\sin\alpha,
~~~~~~~\psi_2\to\psi_1\sin\alpha+\psi_2\cos\alpha,$$ which also
keeps $|\Psi|^2=\psi_{1}^2+\psi_{2}^2$ unchanged. Here $\alpha $
is a constant. On the other hand, in terms of $\Psi=\psi
e^{i\eta}$, where $\psi=|\Psi|$ is the amplitude and $\eta$ is the
phase angle, this transformation is equivalent to
$$\psi\to\psi,~~~~~~~\eta\to\eta+\alpha,$$
which means a phase displacement. According to the well-known
No$\ddot{\rm e}$ther theorem, this symmetry leads to a conserved
charge. In the light of canonical complex scalar field, it is easy
to find that the hessence also has a similar symmetry. One can
verify that the kinetic energy terms
$$\frac{1}{4}\left[(\partial_\mu \Phi)^2+(\partial_\mu \Phi^\ast)^2\right]=
\frac{1}{2}\left(\partial_{\mu}\phi_1\right)^2-\frac{1}{2}\left(\partial_{\mu}\phi_2\right)^2$$
of the hessence is invariant under the transformation
\bea
\phi_1\to\phi_1\cos\alpha-i\phi_2\sin\alpha,\nonumber\\
\phi_2\to-i\phi_1\sin\alpha+\phi_2\cos\alpha,\label{eq13} \eea
which also keeps $\phi_{1}^2-\phi_{2}^2$ unchanged. Then, if the
potential of the hessence $V(\Phi,\Phi^\ast)$ or
$V(\phi_1,\phi_2)$  depends on the quantity $\Phi^2+\Phi^{\ast 2}$
or $\phi_{1}^2-\phi_{2}^2$ only, the Lagrangian density of the
hessence is invariant  under this transformation above. In this
case, the hessence should has a conserved charge. However, we find
that it is unclear to understand the physical meaning of the
transformation Eq.~(\ref{eq13}) in terms of the traditional
formalism of the complex scalar field, i.e. $(\phi_1,\phi_2)$ or
$\Phi=Re^{i\Theta}$. In addition, we find that the equations of
the hessence are very involved in terms of $(R,\Theta)$ while it
is convenient in the case of spintessence. We must find out a new
formalism to describe the new non-canonical complex scalar field,
namely the hessence.

It is suggestive to note that (i) $\phi_{1}^2-\phi_{2}^2=const.$ is a hyperbola
on the $\phi_1$ vs
$\phi_2$ plane, and (ii) by the relations between angular function and hyperbolic
function, one has
\bea
&\sinh z=-i\sin (iz),~~~~~~~\sin z=-i\sinh (iz),\nonumber\\
&\cosh z=\cos (iz),~~~~~~~\cos z=\cosh (iz).\nonumber
\eea
In terms of hyperbolic function, the transformation Eq.~(\ref{eq13}) can be rewritten as
\bea
\phi_1\to\phi_1\cosh (i\alpha)-\phi_2\sinh (i\alpha),\nonumber\\
\phi_2\to-\phi_1\sinh (i\alpha)+\phi_2\cosh (i\alpha).\label{eq14}
\eea
Furthermore, we introduce two new variables $(\phi,\theta)$ to describe the hessence, i.e.
\be{eq15}
\phi_1=\phi\cosh\theta,~~~~~~~\phi_2=\phi\sinh\theta,
\ee
which are defined by
\be{eq16}
\phi^2=\phi_{1}^2-\phi_{2}^2,~~~~~~~\coth\theta=\frac{\phi_1}{\phi_2}.
\ee
And then, the transformation Eq.~(\ref{eq14}) is equivalent to
\be{eq17}
\phi\to\phi,~~~~~~~\theta\to\theta-i\alpha,
\ee
which means an internal ``imaginary motion''. From now on, we will use the new formalism
$(\phi,\theta)$ to describe the new non-canonical complex scalar field. Here, one may see
that the name ``hessence'' arises from the prefix ``h-'' stands for ``hyperbolic'' and the
traditional suffix ``-essence'' for dark energy.

Here, let us have a pause before discussing some physical aspects
of the hessence. Strictly speaking, by the Lagrangian density
Eq.~(\ref{eq11}), the hessence is identified with the quintom
given by Eq.~(\ref{eq6}). However, the potential of the hessence
is imposed to depend only on the quantity $\Phi^2+\Phi^{\ast 2}$
or $\phi_{1}^2-\phi_{2}^2$ or the more convenient $\phi$. In this
sense, the hessence is not equivalent to the quintom model
proposed by Feng, Wang and Zhang~\cite{r10,r14}. The quintom model
of dark energy can be viewed as a realization of dark energy which
makes the transition from $w>-1$ to $w<-1$ or vice versa possible.
To implement this, one may employ two independent real scalar
fields as the case of the toy model proposed in~\cite{r10,r14},
while one also may employ a single field whose potential has some
internal symmetry, e.g. the hessence proposed in this paper.
Secondly, because $\phi_1$ and $\phi_2$ are independent in the
quintom model proposed in~\cite{r10,r14}, one may worry about the
possibility of $\phi^2=\phi_{1}^2-\phi_{2}^2$ becoming negative
when $\phi_{1}^2$ is less than $\phi_{2}^2$. However, we stress
that the hessence cannot be identified with the quintom model
proposed in~\cite{r10,r14} once again. The possibility of $\phi^2$
becomes negative never occurs in the hessence model, since
$\phi_{1}^2\geq\phi_{2}^2$ is ensured by definition, see
Eqs.~(\ref{eq15}) and (\ref{eq16}). On the other hand, we remind
that the equation-of-state parameter $w>-1$ or $w<-1$ depends on
$\dot{\phi}_1$ and $\dot{\phi}_2$ rather than $\phi_1$ and
$\phi_2$ themselves. Thus, we should not worry that the
definitions Eqs.~(\ref{eq15}) and (\ref{eq16}) may ruin the
possibility of $w$ cross the phantom divide $w=-1$. Thirdly, in
the hessence model, the potential is imposed to be the form of
$V(\phi)$, or equivalently, $V(\phi_{1}^2-\phi_{2}^2)$. Except the
very special case of $V(\phi)\sim\phi^2$, the $\phi_1$ and
$\phi_2$ are coupled in general. This is different from the
quintom model studied in~\cite{r10,r14,r13,r21}. Finally, we admit
that the Lagrangian density of the hessence has not been  proposed
in the ordinary particle physics or field theory before,  to our
knowledge. And one may feel that it is difficult to understand the
so-called internal ``imaginary motion'' mentioned above. However,
we argue that they are not the reasons prevent us from the
possiblity of using the novel non-canonical complex scalar field,
i.e. hessence, to describe the dark energy. We think that it is
not strange to use a new field to understand the new and unknown
object: dark energy. On the other hand, old conceptions should not
smother any new idea while it may open a new window to look into
the unknown internal world of the dark energy.


\subsection{Formalism}

Assuming that we have been tolerated to continue, let us restart
our discussion with the action
 \be{eq18}
  S=\int
d^{4}x\sqrt{-g}\left(-\frac{\cal R}{16\pi G}+{\cal L}_h+{\cal
L}_m\right),
 \ee
 where the Lagrangian density of the hessence is
given by
\be{eq19}
 {\cal L}_h=\frac{1}{4}\left[\,(\partial_\mu
\Phi)^2+(\partial_\mu \Phi^\ast)^2\,\right]-U(\Phi^2+ \Phi^{\ast
2})=\frac{1}{2}\left[\,(\partial_\mu \phi)^2-\phi^2 (\partial_\mu
\theta)^2\,\right]-V(\phi).
\ee
  Considering a spatially flat FRW
universe with metric
 \be{eq20}
 ds^2=dt^2-a^2 (t)d{\bf x}^2,
 \ee
where $a(t)$ is the scale factor, from Eqs.~(\ref{eq18}) and
(\ref{eq19}), we obtain the equations of motion for $\phi ({\bf
x},t)$ and $\theta ({\bf x},t)$, namely \be{eq21}
\ddot{\phi}+3H\dot{\phi}-\frac{\nabla^2}{a^2}\phi+\phi\,(\partial_\mu
\theta)^2+V^\prime (\phi)=0, \ee \be{eq22} \phi^2
\ddot{\theta}+(3H\phi^2+2\phi\dot{\phi})\dot{\theta}-\phi^2
\frac{\nabla^2}{a^2}\theta -\frac{2\phi}{a^2}\,\partial_i
\phi\,\partial_i \theta=0, \ee where $H\equiv\dot{a}/a$ is the
Hubble parameter, $\nabla^2\equiv\partial_i \partial_i$, an
overdot and a prime denote the derivatives with respect to cosmic
time $t$ and $\phi$, respectively. If $\phi$ and $\theta$ are
homogeneous, the above equations become \be{eq23}
\ddot{\phi}+3H\dot{\phi}+\phi\dot{\theta}^2+V^\prime (\phi)=0, \ee
\be{eq24} \phi^2
\ddot{\theta}+(2\phi\dot{\phi}+3H\phi^2)\dot{\theta}=0. \ee The
pressure and energy density of the hessence are \be{eq25}
p_h=\frac{1}{2}\left(\dot{\phi}^2-\phi^2
\dot{\theta}^2\right)-V(\phi), \ee \be{eq26}
\rho_h=\frac{1}{2}\left(\dot{\phi}^2-\phi^2
\dot{\theta}^2\right)+V(\phi), \ee respectively. The corresponding
equation-of-state parameter is given by
 \be{eq27}
w=\frac{p_h}{\rho_h}=\frac{\frac{1}{2}\left(\dot{\phi}^2-\phi^2
\dot{\theta}^2\right)-V(\phi)}
{\frac{1}{2}\left(\dot{\phi}^2-\phi^2
\dot{\theta}^2\right)+V(\phi)}.
 \ee
  It is easy to see that
$w\geq-1$ when $\dot{\phi}^2\geq\phi^2 \dot{\theta}^2$ and $w<-1$
when $\dot{\phi}^2<\phi^2 \dot{\theta}^2$. The Friedmann equations
read as \bea &\disp H^2=\frac{8\pi
G}{3}\left[\rho_m+\frac{1}{2}\left(\dot{\phi}^2-\phi^2
\dot{\theta}^2\right)
+V(\phi)\right],\label{eq28}\\
&\disp \frac{\ddot{a}}{a}=-\frac{8\pi
G}{3}\left[\frac{\rho_m}{2}+\left(\dot{\phi}^2- \phi^2
\dot{\theta}^2\right)-V(\phi)\right],\label{eq29} \eea where
$\rho_m$ is the energy density of dust matter.

It is worth noting that if $\dot{\theta}^2\sim 0$, the hessence
reduces to an ordinary quintessence. From
Eqs.~(\ref{eq25})--(\ref{eq29}), we can see that the phantom-like
role is played by the internal motion $\dot{\theta}$. In addition,
Eq.~(\ref{eq24}) implies
\be{eq30} Q=a^3 \phi^2
\dot{\theta}=const.
 \ee
 which is associated with the
total conserved charge within the physical volume. It turns out
\be{eq31} \dot{\theta}=\frac{Q}{a^3 \phi^2}. \ee Substituting into
Eqs.~(\ref{eq23}) and (\ref{eq25})--(\ref{eq29}), we can recast
them as
\be{eq32}
\ddot{\phi}+3H\dot{\phi}+\frac{Q^2}{a^6\phi^3}+V^\prime (\phi)=0,
\ee
\bea
&\disp p_h=\frac{1}{2}\dot{\phi}^2-\frac{Q^2}{2a^6
\phi^2}-V(\phi),~~~~~~~
\rho_h=\frac{1}{2}\dot{\phi}^2-\frac{Q^2}{2a^6 \phi^2}+V(\phi),\nonumber\\
&\disp w=\left.\left[\frac{1}{2}\dot{\phi}^2-\frac{Q^2}{2a^6 \phi^2}-V(\phi)\right]\right/
\left[\frac{1}{2}\dot{\phi}^2-\frac{Q^2}{2a^6 \phi^2}+V(\phi)\right],\label{eq33}
\eea
\bea
&\disp H^2=\frac{8\pi G}{3}\left[\rho_m+\frac{1}{2}\dot{\phi}^2-\frac{Q^2}{2a^6 \phi^2}
+V(\phi)\right],\nonumber\\
&\disp \frac{\ddot{a}}{a}=-\frac{8\pi G}{3}\left[\frac{\rho_m}{2}+\dot{\phi}^2-
\frac{Q^2}{a^6 \phi^2}-V(\phi)\right].\label{eq34}
\eea

From Eq.~(\ref{eq30}), one finds that the sign of the conserved
charge $Q$ is determined by the sign of $\dot{\theta}$. The
conserved charge $Q$ is positive for the case of $\dot{\theta}>0$
while $Q$ is negative for the case of $\dot{\theta}<0$. On the
other hand, it is easy to see that the governing equations, namely
Eqs.~(\ref{eq23}) and (\ref{eq25})--(\ref{eq29}) [or
Eqs.~(\ref{eq32})--(\ref{eq34})] are the same for the cases of
$\dot{\theta}>0$ and $\dot{\theta}<0$, since they depend on
$\dot{\theta}^2$ or $Q^2$ rather than $\dot{\theta}$ or $Q$
themselves. It is possible that there are {\em dark energy and
anti-dark energy} with opposite conserved charges $Q$ in the
universe, just like electron and positron. This new discovery may
have some interesting implications to cosmology. For example, one
may develop some cosmological observations attempting to find the
signals coming  from the annihilation of dark energy and anti-dark
energy. On the other hand, if such observations cannot find out
the anti-dark energy, it means an asymmetry between dark energy
and anti-dark energy, just like the case of baryons and
anti-baryons. Putting these two asymmetries together, it may give
a novel solution to the baryogenesis. Besides, if the conserved
charge of the dark energy (hessence) corresponds to a kind of
long-range force, just like the repulsive force between an
assembly of electrons, they repel each other. Therefore, it is
easy to understand why the dark energy is spatially homogeneous
and they are not clumped to form structures. The novel features of
the hessence are quite interesting for cosmology. We regard these
as a new window to look into the internal world of the mysterious
dark energy. A deeper understanding to dark energy might be
possible through this new window.


\subsection{Dynamics}

Obviously, from Eq.~(\ref{eq33}), the equation-of-state parameter $w\geq -1$ when
$\dot{\phi}^2\geq Q^2/(a^6 \phi^2)$ while $w<-1$ when $\dot{\phi}^2< Q^2/(a^6 \phi^2)$.
The transition occurs when $\dot{\phi}^2=Q^2/(a^6 \phi^2)$.

\begin{center}
\begin{figure}[htp]
\includegraphics[width=0.49\textwidth]{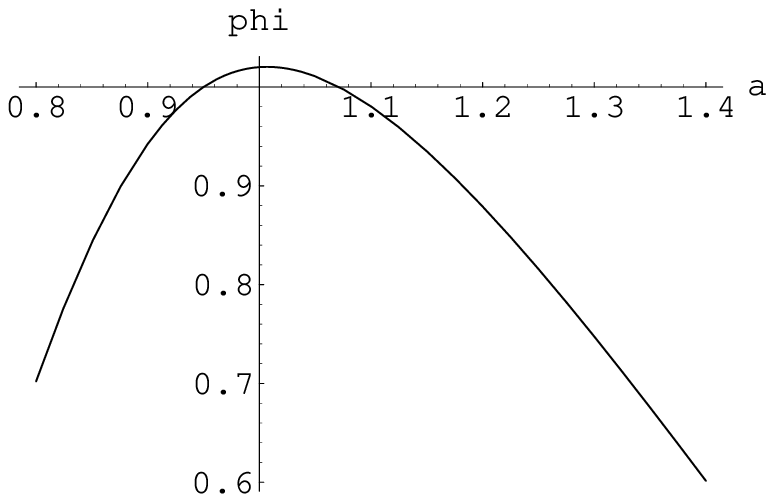}\hfill
\includegraphics[width=0.49\textwidth]{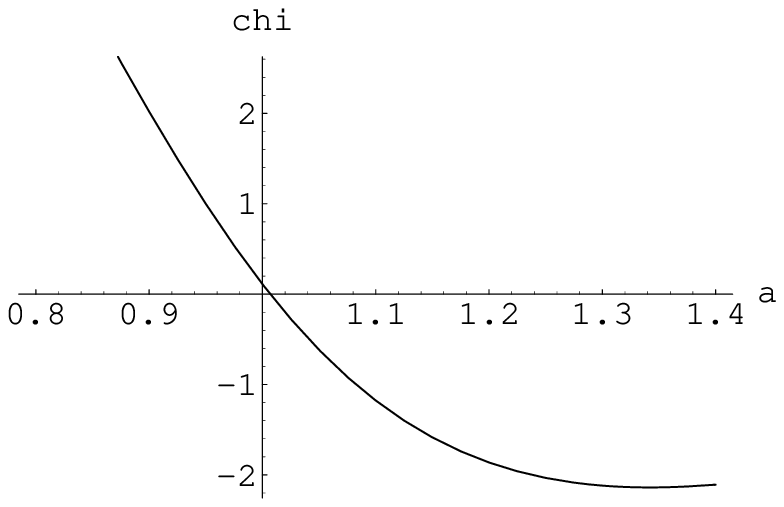}\\
\centering
\includegraphics[width=0.5\textwidth]{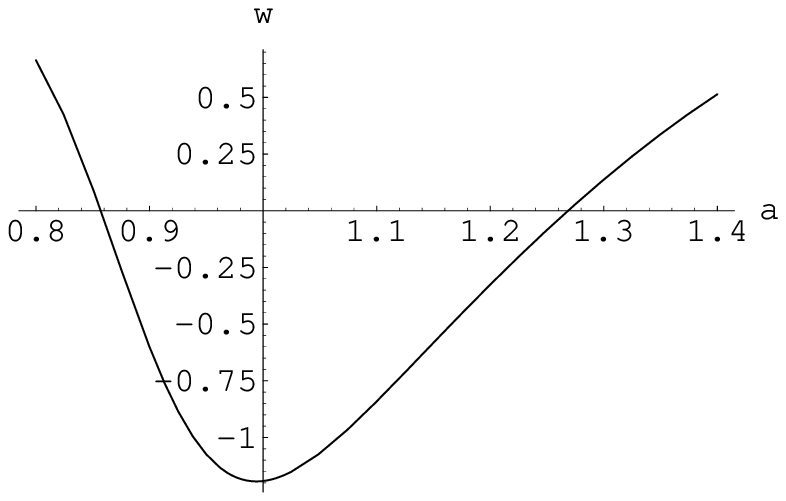}
\caption{\label{fig1} The numerical plots of $\phi$, $\chi\equiv\dot{\phi}$ and the
equation-of-state parameter $w$ versus scale factor $a$ for the $V(\phi)=\lambda\phi^4$ 
potential. We choose the demonstrative parameters as $\Omega_{m0}\equiv\rho_{m0}/(3H_0^2)=0.3$, 
$\lambda=5.0$, $Q=1.0$. We set the scale factor $a_0=1$ and the unit $8\pi G=1$. The 
equation-of-state parameter $w$ goes beyond $-1$ at $a=0.95$.}
\end{figure}
\end{center}

In fact, it is difficult to obtain the analytic solutions for the equation of motion of
hessence. To see the dynamics of the hessence, we have to adopt the numerical
approach. To this end, we recast Eq.~(\ref{eq32}) and the first Friedmann equation
Eq.~(\ref{eq34}) as following first-order differential equations with respect to the
scale factor $a$
$$\frac{d\phi}{d a}=\frac{\chi}{aH},$$
$$\frac{d\chi}{d a}=\frac{-1}{aH}
\left[3H\chi+\frac{Q^2}{a^6 \phi^3}+V^\prime (\phi)\right],$$
and
$$H^2=\frac{1}{3}\left[\rho_m+\frac{1}{2}\chi^2-\frac{Q^2}{2a^6 \phi^2}
+V(\phi)\right],$$
where $\chi\equiv\dot{\phi}$. For simplicity, we set the unit $8\pi G=1$ in this
subsection. The equation-of-state parameter is given by
$$w=\left.\left[\frac{1}{2}\chi^2-\frac{Q^2}{2a^6 \phi^2}-V(\phi)\right]\right/
\left[\frac{1}{2}\chi^2-\frac{Q^2}{2a^6 \phi^2}+V(\phi)\right].$$
We consider the case of minimal couple between hessence and dust matter, thus
$$\rho_m=\rho_{m0}a^{-3},$$
where the subscript ``0'' indicates the present value of corresponding quantity. To
be definite, we take the potential
$$V(\phi)=\lambda\phi^4$$
for example, while the case of $V(\phi)\sim\phi^2$ is trival. We show the numerical
result in Fig.~\ref{fig1}.

It is obvious that the equation-of-state parameter $w$ can cross
the phantom divide $w=-1$ indeed. Although one may find that
$w<-1$ is transient in the case of $\phi^4$ potential, it is worth
noting that the behavior of $w$ depends heavily on the form of the
potential $V(\phi)$. The case of $\phi^4$ potential presented here
is only for a naive demonstration. One can build a more realistic
 hessence dark energy model to fit the observation data by
choosing a proper potential.


\section{Free of the Q-ball formation}
The formation of  Q-balls, is very generic for a complex field
(see~\cite{r20} for example). A Q-ball is a kind of nontopological
soliton whose stability is guaranteed by some conserved charge
Q~\cite{r19}. In the case of spintessence~\cite{r15,r16,r17} which
is  a canonical complex scalar field mentioned above, it is
difficult to avoid the Q-ball formation~\cite{r17,r15}. Except in
some special cases of spintessence with an unnatural potential,
the fluctuations grow exponentially and go nonlinearly to form
Q-balls. Once the Q-balls are formed, they will act as  (dark)
matter whose energy density decreases as $a^{-3}$. As for the
late-time fate of the Q-balls, it depends on the shape of the
potential, and they can be stable to be the dark matter, or decay
into other particles like radiations whose energy density
decreases as $a^{-4}$. Therefore, the spintessence cannot be a
viable candidate of the dark energy (as pointed out in~\cite{r15},
however, the spintessence may be a good candidate of cold dark
matter). As a non-canonical complex scalar field, the hessence
faces a similar situation. Fortunately,  note that the terms
$\dot{\theta}^2$ in the equations of spintessence correspond to
$-\dot{\theta}^2$ in our hessence case, and that this term is
crucial in the criterion of the Q-ball formation~\cite{r17}. Thus,
we are optimistic to expect that the hessence can avoid this
difficulty by the help of the negative sign. It turns out it is
true.

We now consider the growth of perturbations in the hessence.
Following~\cite{r17}, we assume that the gravity effect is weak,
which is a good approximation here. Thus, we do not consider the
metric perturbation arising from the fluctuations in the hessence
and surrounding matter. Substituting $\phi({\bf
x},t)=\phi(t)+\delta\phi({\bf x},t)$ and $\theta({\bf
x},t)=\theta(t)+ \delta\theta({\bf x},t)$ into the equations of
motion of $\phi$ and $\theta$, i.e. Eqs.~(\ref{eq21}) and
(\ref{eq22}), and linearizing the resulting equations, we obtain
\be{eq35}
  \ddot{\delta\phi}+3H\dot{\delta\phi}+2\phi\dot{\theta}\dot{\delta\theta}+\dot{\theta}^2
\delta\phi+
V^{\prime\prime}(\phi)\delta\phi-\frac{1}{a^2}\nabla^2\delta\phi=0,
\ee
  \be{eq36}
    \phi^2 \ddot{\delta\theta}+3H\phi^2
\dot{\delta\theta}+2\phi\left(\dot{\phi}\dot{\delta\theta}+
\dot{\theta}\dot{\delta\phi}\right)-2\dot{\phi}\dot{\theta}\delta\phi-
\frac{\phi^2}{a^2}\nabla^2\delta\theta=0,
 \ee
  for fluctuations. We seek
for the solutions in the form \be{eq37} \delta\phi=\delta\phi_0\,
e^{\omega t+i{\bf k}\cdot{\bf x}},~~~~~~~
\delta\theta=\delta\theta_0\, e^{\omega t+i{\bf k}\cdot{\bf x}}.
\ee If $\omega$ is real and positive, these fluctuations grow
exponentially and go nonlinearly to form Q-balls. Substituting
Eq.~(\ref{eq37}) into Eqs.~(\ref{eq35}) and (\ref{eq36}), one has
\be{eq38}
\left[\omega^2+3H\omega+\dot{\theta}^2+V^{\prime\prime}(\phi)+\frac{k^2}{a^2}
\right]\delta\phi_0+ 2\omega\phi\dot{\theta}\delta\theta_0=0, \ee
 \be{eq39}
2\dot{\theta}\left(\phi\omega-\dot{\phi}\right)\delta\phi_0+\left(\phi^2
\omega^2+3H\phi^2 \omega+ 2\phi\dot{\phi}\omega+\phi^2
\frac{k^2}{a^2}\right)\delta\theta_0=0.
\ee
 The condition for
nontrivial $\delta\phi_0$ and $\delta\theta_0$ is given by
\be{eq40}
\left[\omega^2+3H\omega+\dot{\theta}^2+V^{\prime\prime}(\phi)+\frac{k^2}{a^2}\right]\times
\left(\phi^2 \omega^2+3H\phi^2 \omega+2\phi\dot{\phi}\omega+\phi^2
\frac{k^2}{a^2}\right) =4\omega\phi\dot{\theta}^2
\left(\phi\omega-\dot{\phi}\right).
 \ee
  Assuming that cosmological
expansion effect is negligible,  we pay special attention to the
case with a rapidly varying $\theta$ since the hessence reduces to
the quintessence as $\dot{\theta}^2\sim 0$ mentioned above, namely
$H\sim 0$, and $\phi\sim const.$. In this case, the condition Eq.~(\ref{eq40}) becomes
\be{eq41}
\omega^4+\left(2\frac{k^2}{a^2}+V^{\prime\prime}-3\dot{\theta}^2\right)
\omega^2+\left(\frac{k^2}{a^2}+
V^{\prime\prime}+\dot{\theta}^2\right)\frac{k^2}{a^2}=0. \ee
 We find that the Jeans wavenumber $k_J$ at which $\omega^2=0$ is
given by
\be{eq42}
\frac{k_{J}^2}{a^2}=-\dot{\theta}^2-V^{\prime\prime}.
 \ee
 If the
Jeans wavenumber exists, the instability band is
 \be{eq43}
0<\frac{k^2}{a^2}<\frac{k_{J}^2}{a^2}. \ee
  However, it is easy to
see that if \be{eq44} \dot{\theta}^2+V^{\prime\prime}\geq 0, \ee
the instability band does not exist. Then the Q-balls cannot be
formed. This condition Eq.~(\ref{eq44}) is not difficult to be
satisfied for many potentials, such as $V(\phi)=V_0
(\phi_{0}/\phi)^n$, $V(\phi)=V_0 [\exp (\phi_{0}/\phi)-1]$,
$V(\phi)=V_0 \exp (-\lambda\phi)$ etc.~\cite{r18}. It is easy to
see that the negative sign in front of $\dot{\theta}^2$ in
Eq.~(\ref{eq42}) is crucial to prevent from the Q-ball
formation~\cite{r17,r15}. On the other hand, $w<-1/3$ does not
restrict $V^{\prime\prime}<0$ for our case considered here, unlike
the case of spintessence. After all, as illustrated
in~\cite{r15,r18}, the essential information about the behavior of
perturbations for the hessence should be still valid in the
full-blown relativistic analysis.


\section{Hessence and Chaplygin gas}

In~\cite{r24}, the so-called (generalized) Chaplygin gas was
studied as an alternative to quintessence. Actually, the
inhomogeneous (generalized) Chaplygin gas may be an unification of
dark energy and dark matter. Furthermore, it was found that the
(generalized) Chaplygin gas can arise from brane, quintessence,
tachyon etc.. In addition, the (generalized) Chaplygin gas can be
described by a formalism of canonical complex scalar field also~\cite{r24}.
Thus, it is interesting to find the possible relation between the
hessence and the Chaplygin gas. In this section, we will show
that, by choosing a proper potential, the hessence  can be
described by a Chaplygin gas at late times.

The Chaplygin gas is an exotic fluid described by the equation of
state \be{eq45} p=-\frac{A}{\rho}, \ee where $A$ is a positive
constant. It can be generalized to the so-called generalized
Chaplygin gas whose equation of state is given by
\be{eq46}
p=-\frac{A}{\rho^\beta},
 \ee
 where $\beta$ is also a positive
constant. Using the relativistic energy-momentum conservation
equation
\be{eq47}
 \dot{\rho}+3H(p+\rho)=0,
 \ee one has
\be{eq48}
\rho=\left[A-\frac{B}{a^{3(1+\beta)}}\right]^{1/(1+\beta)},
 \ee
where $B$ is an integration constant. To find out the possible
relation between the hessence and the Chaplygin gas, we set
\be{eq49}
p_h=-\frac{A}{\rho_{h}^\beta},
 \ee
 where $\rho_h$ is assumed to have the form Eq.~(\ref{eq48}).
 From Eq.~(\ref{eq33}), we get \bea
&\disp 2V(\phi)=\rho_h-p_h=\rho_h+\frac{A}{\rho_{h}^\beta},\label{eq50}\\
&\disp \dot{\phi}^2-\frac{Q^2}{a^6
\phi^2}=\rho_h+p_h=\rho_h-\frac{A}{\rho_{h}^\beta}.\label{eq51}
\eea
 Considering a hessence-dominated universe, the Friedmann
equation reads \be{eq52}
\left(\frac{\dot{a}}{a}\right)^2=H^2=\frac{8\pi G}{3}\rho_h. \ee
Substituting $\dot{\phi}=\dot{a}(d\phi/da)$ and Eqs.~(\ref{eq48}),
(\ref{eq52}) into Eq.~(\ref{eq51}), we can recast it as a
differential equation of $\phi$ with respect to $a$. In principle,
we can solve it and obtain the $\phi(a)$. Then we get $V(\phi)$
from Eq.~(\ref{eq50}) by using $\phi(a)$ and Eq.~(\ref{eq48}).
However, we find that it is difficult to solve out the $\phi(a)$
for $Q\not=0$ and/or $\beta\not=1$ case. As mentioned above, $Q=0$
case is trivial since the hessence reduces to the quintessence. To
find a sensible solution, we consider the case with a  rapidly
varying $\theta $, namely,
 \be{eq53}
\dot{\phi}^2\ll\frac{Q^2}{a^6 \phi^2},
 \ee
  and  $\beta=1$ [see
Eq.~(\ref{eq45})]. In this case, Eqs.~(\ref{eq48}) and
(\ref{eq51}) become \bea
&\disp\rho_h=\sqrt{A-\frac{B}{a^6}},\label{eq54}\\
&\disp-\frac{Q^2}{a^6
\phi^2}=\rho_h-\frac{A}{\rho_{h}}.\label{eq55} \eea It is easy to
get $\rho_h(\phi)$ and $a(\phi)$ as 
\be{eq56}
\rho_h=\frac{B\phi^2}{Q^2},~~~~~~~a=\left(\frac{BQ^4}{AQ^4-B^2\phi^4}\right)^{1/6}. 
\ee
It is worth noting that to get a real and positive
scale factor $a$ and positive energy density, the
 constant $B$ must satisfy
$B>0$. In addition, one can see from Eq.~(\ref{eq54}) that it is valid
only when $A a^6 \ge B$.
 Further  one can get
$\phi(a)$ as \be{eq57}
\phi=\pm\left(A-\frac{B}{a^6}\right)^{1/4}\frac{Q}{\sqrt{B}}. \ee
From Eq.~(\ref{eq50}), we obtain the corresponding potential
\be{eq58}
 V(\phi)=\frac{B\phi^2}{2Q^2}+\frac{AQ^2}{2B\phi^2},
 \ee
which is a quite simple form.

Finally, we will show the compatibility of the calculations above.
By using $\dot{\phi}= \dot{a}(d\phi/da)$ and Eqs.~(\ref{eq52}),
(\ref{eq54}), (\ref{eq57}), we have \be{eq59}
\dot{\phi}^2=\frac{6\pi GBQ^2}{a^{12}\left(A-B/a^6\right)}. \ee On
the other hand, from Eq.~(\ref{eq57}), \be{eq60} \frac{Q^2}{a^6
\phi^2}=\frac{B}{a^6 \sqrt{A-B/a^6}}. \ee Obviously, the condition
Eq.~(\ref{eq53}) is satisfied provided that the scale factor $a$
is large. That is, it is at late times.


\section{Summary and discussions}
In summary,  we propose a non-canonical complex scalar field,
which we dub ``hessence'', to implement the concept of quintom
dark energy whose equation-of-state parameter $w$ can cross the
phantom divide $w=-1$. In the hessence model, the dark energy is
described by a single field with an internal degree of freedom
rather than two independent real scalar fields. Furthermore, the
hessence is imposed an internal symmetry and then, it has a
conserved charge. We develope a new formalism to describe the new
non-canonical complex scalar field, i.e. hessence. We find that in
the hessence model, the phantom-like role is played by the
internal motion. We regard this hessence model as a new window to
look into the unknown internal world of the mysterious dark
energy. In addition, we show that the hessence can avoid the
difficulty of the Q-ball formation which gives trouble to the
spintessence. Furthermore, we find that, by choosing a proper
potential, the hessence  can be described  by a Chaplygin gas at
late times.

 Although the cosmological
evolution of the quintom model proposed in~\cite{r10} was studied
by Guo {\it et al.}~\cite{r14}, we find that it is still
interesting to investigate the cosmological evolution of the
hessence. In fact, the authors of \cite{r14} only considered the
special case whose potential $V(\phi_1,\phi_2)$ can be decomposed
into $V(\phi_1)+V(\phi_2)$, namely the case in which there is no
direct coupling between $\phi_1$ and $\phi_2$. However, in the
hessence model, the potential is imposed to be the form of
$V(\phi)$, or equivalently, $V(\phi_{1}^2-\phi_{2}^2)$. Except the
very special case of $V(\phi)\sim\phi^2$, the $\phi_1$ and
$\phi_2$ are coupled in general. Therefore it is of interest to
further investigate the cosmological evolution of the hessence.
Besides, there are some interesting open questions, such as

\begin{itemize}

\item Can the hessence arise from a more fundamental theory, such as string/M theory
or braneworld model?

\item How to construct the quantum field theory for the hessence? As a non-canonical
complex scalar field, its Lagrangian density never appeared  in
the ordinary particle physics and  quantum field theory,  to our
knowledge.

\item What role may the conserved charge of the hessence play in cosmology? Can we
imagine the novel possibility of ``dark energy and anti-dark energy'' and so on?

\item Whether is the hessence  stable in the quantum level? While the phantom is not
stable in this level~\cite{r7,r32}, one may worry about the hessence since it contains
a phantom-like ingredient.

\end{itemize}


\section*{Acknowledgments}
HW is grateful to Prof. Xin-Min Zhang and Bo Feng, Xiao-Jun Bi,
Ming-Zhe Li, Hong Li, Xin Zhang of IHEP for suggestive
discussions. He also thanks  Xin-Qiang Li, Zong-Kuan Guo, Qi Guo,
Hong-Sheng Zhang, Hui Li, Xun Su, Fei Wang,
Fu-Rong Yin and Wei-Shui Xu of ITP for helpful discussions. This
work was supported by a grant from Chinese Academy of Sciences,
grants from NSFC, China (No. 10325525 and No. 90403029), and a
grant from the Ministry of Science and Technology of China (No.
TG1999075401).


\section*{Note added}
After our paper was submitted, some papers concerning this issue
appeared in the arXiv preprint~\cite{r25,r26,r27,r28,r29,r30,r36}. In
particular, the data-fit of Ref.~\cite{r30} shows that the 
SNe Ia Gold dataset favors the equation-of-state parameter $w$ cross
the phantom divide $w=-1$. In Ref.~\cite{r25}, the quintom model
with the special case of interaction
$V_{int}(\phi_1,\phi_2)\sim\sqrt{V(\phi_1)V(\phi_2)}$ has been
studied. On the other hand, a novel single real scalar field model
with $w$ crossing $-1$ has been proposed in Ref.~\cite{r28},
whose Lagrangian density contains second-order differential term
of the scalar field (to break the fourth condition of the no-go
theorem~\cite{r12} mentioned above). This kind of Lagrangian
density can be used to drive the so-called B-inflation~\cite{r31}
also. Furthermore, it was shown in Ref.~\cite{r36} that $w$ crossing $-1$ 
is possible without introducing any phantom component in a Gauss-Bonnet 
braneworld with induced gravity.


\end{document}